\documentstyle[editedvolume,namedreferences,epsfig]{crckapb}

\begin{opening}
\title{SHAPING PLANETARY NEBULAE: is it different for [WR] stars?}

\author{GARRELT MELLEMA} \institute{Stockholm Observatory\\ SE--133~36
Saltsj{\"o}baden, Sweden}

\end{opening}

\runningtitle{SHAPING [WR] PLANETARY NEBULAE}

\begin{document}


\section{Introduction}

The Wolf--Rayet phenomenon is a stellar phenomenon. This means that its
length scale is essentially that of the star, i.e.\ some $10^9$~cm. A typical
nebular size is $10^{14}$~cm or higher, which immediately illustrates that it
may not be easy to make the connection between the star and the nebulae and
observations confirm this. In this paper I will try to discuss some ways in
which the nebular properties could be changed by the presence of a [WR]
central star.

The star's {\it actio in distans}\/ is through two means
\begin{enumerate}
\item{Photons}

The ionizing photons from the star heat and ionize the gas. This can have
non--trivial consequences for the shaping of the PN, as shown by Marten \&
Sch{\"o}nberner (1991), Mellema~(1994), and Mellema~(1995). However, although
the spectral energy distribution of a [WR] star is quite different from that
of a `normal' central star of a PN (CSPN), I do not expect this difference to
cause any big differences in the hydrodynamic structure of the PN. The
detailed ionization structure can be quite different, as illustrated by the
photo--ionization modelling of Crowther et al.~(1999) for the nebulae M1--67,
but the ways in which the onset of ionization modifies the density of the
circumstellar matter through large or small scale ionization fronts will
probably not be very different.
\item{Stellar wind}

The stellar wind differs in two aspects from a normal CSPN wind
\begin{description}
\item [a) Mass loss rate.]

The stellar wind from a [WR] star is generally a factor 10--100 more massive
than from a normal CSPN, see for instance Leuenhagen et al.~(1998). This
means that the wind luminosity (${1\over 2} \dot{M} v^2$) and its momentum
($\dot{M}v$) will be a factor 10--100 higher. This should have an effect on
the formation of the PN.

\item [b) Abundances.]

The wind from a typical [WC] star consists for approximately 0\% of H, 50\%
of He, 40\% of C and 10\% of O, abundances radically different from the usual
cosmic or solar abundances. As a wind with these abundances starts
interacting with the environment its radiative cooling behaviour will be very
different from the standard case. Also this will have an effect on the
shaping of the PN.
\end{description}
\end{enumerate}

Apart from what we observe {\it now}\/ the mass loss history of the star is
also important for understanding the shaping. Did the star become a [WR]
directly after the AGB or is it the result of a late or very late He-shell
flash? In the first case, was the mass loss during the AGB different for the
[WR] stars or did the event which turned the star into a [WR] star influence
its mass loss? Which role does binarity play in [WR] systems, and are there
accretion disks?

These are all questions to which the answer is difficult to give since even
for normal PNe they are not always well known. The standard model to explain
the formation of PNe is the Interacting Stellar Winds [ISW] model or
sometimes called Generalized ISW [GISW] model. In this model the nebulae is
formed from the interaction between the slow AGB wind and the faster
post--AGB wind. The usual assumption (confirmed by observations) is that the
AGB wind has an aspherical density distribution (disk or torus), which leads
to the formation of an aspherical PNe. The ISW model goes back to the late
70-ies (Kwok et al.~1978).

In this paper I will discuss how the formation of a PN can be different for
[WR] central stars, taking the ISW model as the basis.  However, recently the
ISW model has received some criticism which may be good to list here, even
though we do not know how relevant these problems are for the PNe around [WR]
stars.
\begin{itemize}
\item Although aspherically distributed circumstellar matter is commonly
observed in PNe, it is not around AGB stars. The puzzle then is how and when
did the mass loss turn from more or less spherical to aspherical. The ISW
model does not address this point, it just assumes the asphericity. A
complete model should include a mechanism for producing aspherical mass
loss. Most observational evidence indicates that the transition to aspherical
mass loss happens right at the end of the AGB.
\item Jets and other collimated outflow phenomena are observed in some
PNe. Although the ISW model can produce some degree of collimation, it seems
unlikely at this point that the observed jets can be explained by the ISW,
especially since they often show point-symmetry.
\item Point--symmetry is not only shown by jets and collimated outflow
phenomena, but even by whole nebulae. This is indicative of changes in
direction of the symmetry axis of the system. Again, this seems difficult to
establish in the case of the ISW model, since this would require a warped
density distribution in the AGB wind.
\item Young PNe, and many pre-PNe show aspherical morphologies, often with
very peculiar shapes (see for example Sahai \& Trauger 1998; Ueta et
al.~1999). Since the post--AGB wind only becomes energetic for high effective
temperatures, it is unclear how the ISW model can produce the observed
pre-PNe and young PNe. Also, some of the shapes of the young PNe seem to be
in conflict with the ISW model; we see objects with several symmetry axes,
wave function--like morphologies, etc.
\end{itemize}

To deal with these difficulties some alternatives and/or additions to the ISW
model have been proposed. The role of companions (stellar or substellar) for
establishing aspherical mass loss has become generally accepted, albeit more
for lack of alternatives than solid observational proof. See Soker (1998) and
references therein for ways in which gravitational interaction with a
companion can lead to aspherical mass loss. To explain jets and jet-like
phenomena accretion disks have been proposed, possibly around a companion
object. Sahai \& Trauger (1998) even suggested that a short jet phase is
responsible for moulding the initially spherical AGB wind into a torus like
density distribution. A third alternative is the existence of strong magnetic
fields in the post--AGB wind which through hoop stresses can lead to the
formation of aspherical PNe (Chevalier \& Luo 1995). These `MHD models' can
possibly also explain jets and point-symmetry (Garcia--Segura et al.~1999).

Thus, the caveat is that the ISW model may not be the whole story, but since
much remains unclear, I will in this paper still concentrate on the ISW
model.

\section{Observations}

The observations of PNe around [WC] stars are reviewed by G{\'o}rny in this
volume. The summary is that there are few to no differences. The only
differences to be noted are
\begin{itemize}
\item The average expansion velocity of the ionized material is observed to
be somewhat higher than in the case of `normal' PNe (G{\'o}rny \&
Stasi{\'n}ska 1995).
\item The line shapes in WR-PNe seem to require a higher value for the
turbulent velocity component than in normal PNe (Ge{\c s}icki \& Acker~1996).
\end{itemize}
As we will see in the following sections both these effects can be understood
from the differences between normal fast winds and those from [WR] stars.

The most puzzling aspect of the observations is the lack of correlation
between [WR] stars and certain types of PN morphologies. All types of
morphology (elliptical, bipolar, attached shells, etc.) are found around [WR]
stars in approximately the same fractions as in normal PNe. Somehow one would
expect the morphology of a PN to be dependent on the mass loss history of
the star. For a [WR] star this must have been different since the entire H
envelope was lost. Reversely, it has been shown that bipolar PNe are
associated with more massive progenitors, and one would expect that the mass
of the star is important to determine whether it becomes a [WR] star or
not. Still there seems to be the normal fraction of bipolar PNe around [WR]
stars.

The conclusion thus seems to be that the process which determines the shapes
of PNe, i.e.\ the start of aspherical mass loss, is unrelated to the process
which turns the star in a [WR]--type star, or at least is not influenced much
by the way in which the star loses its envelope.

This is a rather amazing conclusion. If one for example considers a common
envelope scenario, in which the entire envelope is ejected, one would think
that the case in which all of the H--envelope is removed is more extreme than
the one in which only part of the H-envelope is removed, which would lead to
different morphologies of the subsequent PN. But the observations show that
this is not the case.

This behaviour can be used as a test for proposed mechanisms to introduce
asphericity in the AGB/post-AGB system. For example one might argue that
mixing will be stronger in the case of faster rotation, and therefore some
relation between nebulae around H-poor central stars and more extreme
morphologies should be present. Since there is no such relation, rotation is
a less likely mechanism to introduce asphericity.

Although useful in principle, the application of this test is complicated by
the fact that the real mechanism for producing H-poor central stars is not
fully understood. All the proposed mechanisms centre around thermal pulses
(either a final pulse on the AGB or a late pulse in the post-AGB phase, see
the contributions of Bl{\"o}cker and Herwig in this volume), which play a
marginal role in most of the proposed mechanisms for aspherical mass loss.

Also, the set of [WR]--PNe has not been studied in much detail. A more
thorough investigation of image and kinematic data of individual nebulae may
still reveal some differences, gone unnoticed when using catalogue data. A
more detailed study of a sample of [WR]--PNe should be done before the above
test becomes really hard, but such a study would be worth it.

\section{Massive stellar winds}

The typical fast wind of a normal CSPN has a mass loss rate in the range
$10^{-9}$---$10^{-7}$~M$_\odot$~yr$^{-1}$, whereas the for the [WR] stars the
reported rates are $10^{-7}$---$10^{-5}$~M$_\odot$~yr$^{-1}$, roughly a
factor 100 higher. The wind velocities do not differ much, which is not
surprising since for radiatively driven winds the terminal wind velocity is
of order the escape velocity from the surface of the star.

The result of a more massive stellar wind is that the momentum ($\dot{M}v$)
and energy (${1 \over 2}\dot{M}v^2$) input is a 100 times larger. The main
effect of this is an increase in the expansion velocity of the nebula.

\begin{figure}
\epsfig{figure=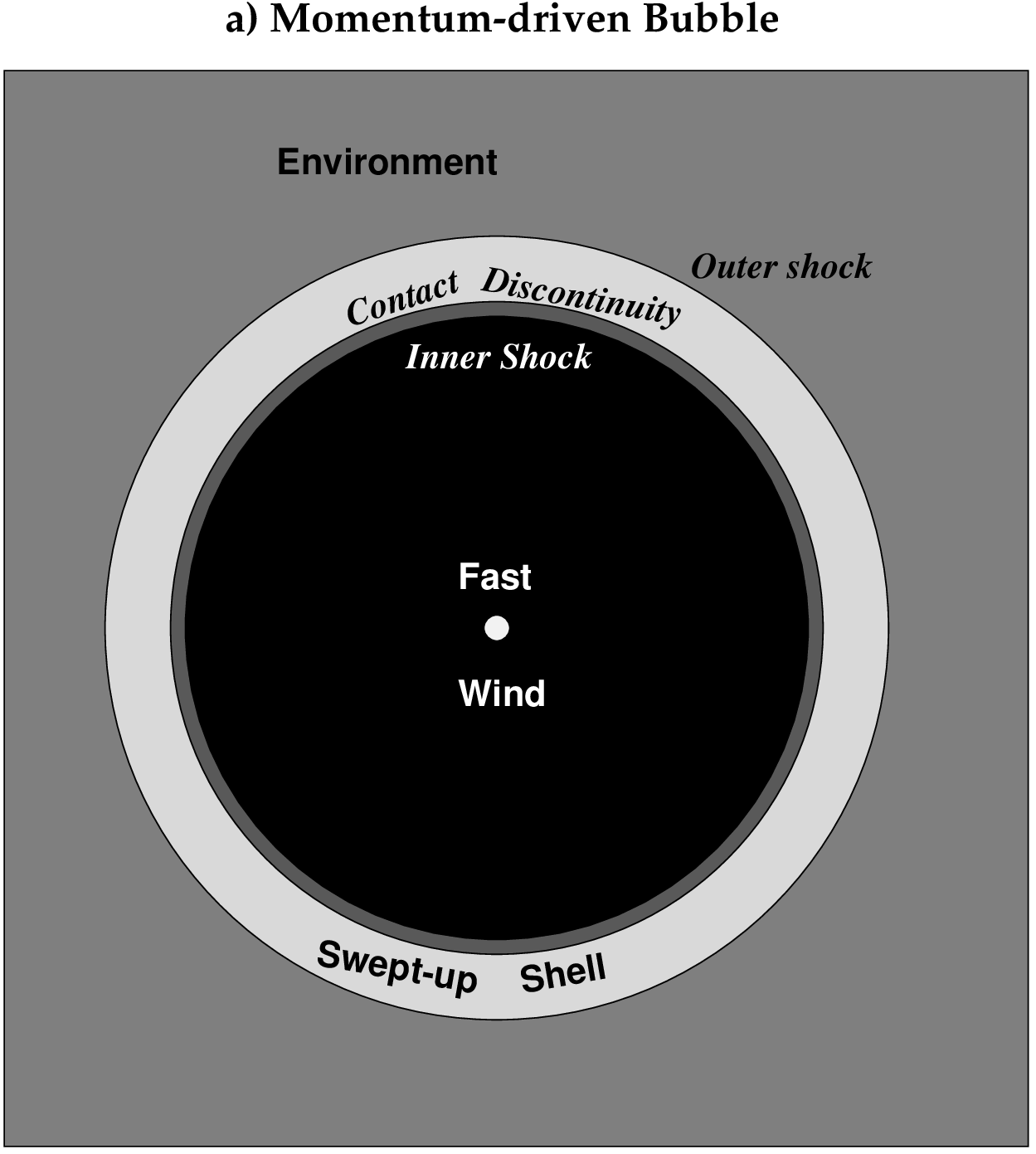,width=7cm,angle=-90}\hfill\epsfig{figure=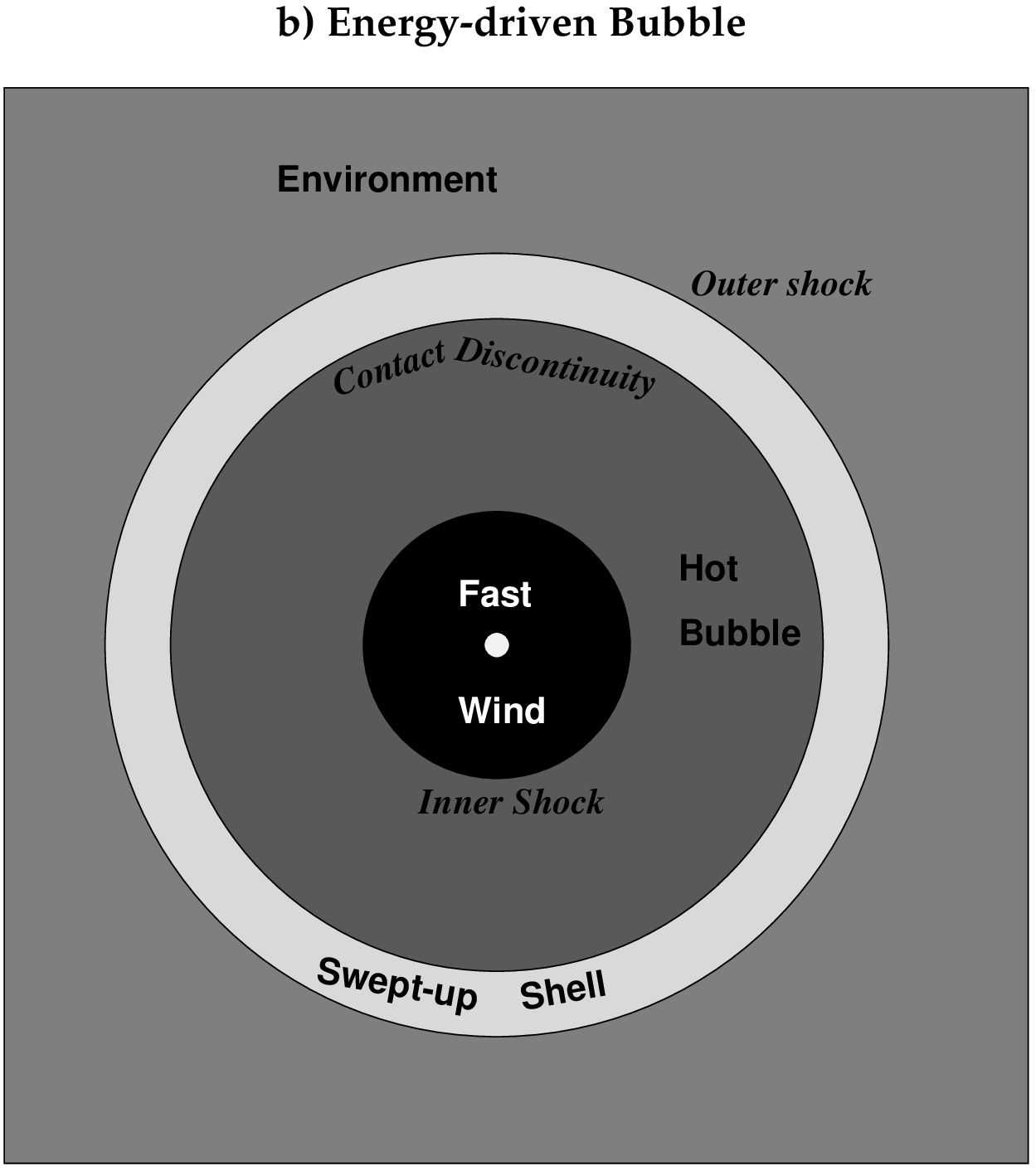,width=7cm,angle=-90}
\caption{Structure of momentum--driven and energy--driven wind bubbles}
\end{figure}
Nebulae formed by a stellar wind come in two types, depending on how well the
stellar wind cools when it is shocked (see e.g.\ Lamers \& Cassinelli 1999,
Ch.~12.3). If this cooling is efficient enough to radiate away all the energy
injected by the stellar wind, the wind--driven bubble is said to be
`momentum--driven'. The structure of the bubble is as shown in Fig.~1a, the
stellar wind fills the entire volume of the bubble and there is a thin cooling
zone separating the stellar wind from the material swept up from the
environment. In this case it is the ram pressure of the stellar wind which
sweeps up a bubble.

If the cooling of the stellar wind is inefficient, a volume of hot, shocked
fast wind material forms and can fill most of the volume of the bubble, with
an inner shock lying relatively close to the star, see Fig.~1b. This type of
bubble is said to be `energy--driven', it is the thermal pressure of the
volume of hot shocked fast wind material which pushes the nebula into the
environment.

There are of course intermediate cases, but in general the division between
the two holds.

Kahn (1983) showed that for an energy--driven bubble the expansion velocity
can be approximated by
\begin{equation}
  v_{\rm exp}=\lambda v_{\rm slow}
\end{equation}
where $\lambda$ is given by the solution of the cubic equation
\begin{equation}
  \lambda(\lambda - 1)^2 = {2 \over 3} {\dot{M}_{\rm fast} v_{\rm fast}^2
  \over \dot{M}_{\rm slow} v_{\rm slow}^2}\,,
\end{equation}
which depends on the ratio of the two wind luminosities (${1 \over 2}\dot{M}
v^2$). An alternative solution was presented by Koo \& McKee (1992). They
derive
\begin{equation}
 v_{\rm exp}= \left(1+ \left ( {2\pi \over 3} \Gamma_{\rm
 rad}\xi{\dot{M}_{\rm fast} v_{\rm fast}^2 \over \dot{M}_{\rm slow} v_{\rm
 slow}^2}\right)^{1/3}\right)v_{\rm slow}\,,
\end{equation}
in which $\Gamma_{\rm rad}$ is the fraction of the energy injected by the
stellar wind (in the case of no radiative losses equal to 1), and $\xi$ a
numerical constant of order unity (whose value is not given by the
authors). These two solutions are equivalent.

For the momentum--driven case Kahn \& Breitschwerdt (1990) found that
\begin{equation}
  v_{\rm exp}= \left(1+\left({\dot{M}_{\rm fast} v_{\rm fast} \over
  \dot{M}_{\rm slow} v_{\rm slow}}\right)^{1/2}\right)v_{\rm slow}\,,
\end{equation}
which depends on the ratio of the two wind momentum rates ($\dot{M} v$).

Fig.~2 shows a plot of the expansion velocity for both cases as a function of
$\dot{M}_{\rm fast}$ for fixed $\dot{M}_{\rm slow}$, $v_{\rm slow}$, and
$v_{\rm fast}$. One sees that the expansion velocity increases as a function
of fast wind mass loss rate, but that the difference becomes large only for
very high mass loss rates. For the chosen parameters the effect is stronger
for the energy--driven case.

\begin{figure}
\epsfig{figure=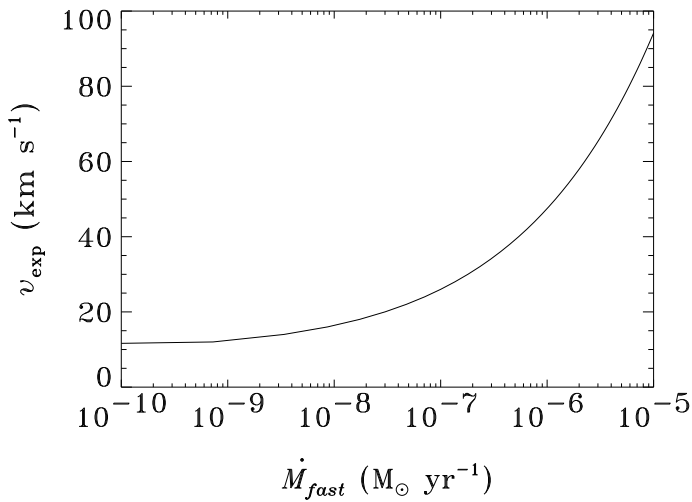,width=10cm}
\caption{Expansion velocity of a spherical wind bubble for different values
of the fast wind mass loss rate.}
\end{figure}

Figure~2 should not be overinterpreted. The spread in mass loss rates and
velocities in a sample of PNe will make the correlation less clear,
especially since aspherical nebulae will have different expansion velocities
in different directions. Also, Fig.~2 neglects any evolution of the winds; we
know that mass loss rates and wind velocities are changing throughout the
post--AGB phase but in calculating the expansion velocities the fast wind
properties are assumed to be constant (the actual assumptions are that
$(\dot{M}v)_{\rm fast}$ is constant in time for the momentum--driven case and
$(\dot{M}v^2)_{\rm fast}$ for the energy--driven case). Still the conclusion
is that we should not be surprised to find an on average higher expansion
velocity for the nebulae around [WR] stars.

\section{Radiative Cooling}

As was outlined above the efficiency of the radiative cooling in the shocked
fast wind determines the character of the wind--driven bubble. Since the
cooling processes are mostly two body interactions, the cooling rate is
proportional to the square of the particle density, $n^2$. The dependence on
the temperature is much stronger. For a gas of cosmic abundances, the cooling
is very strong between temperatures of $10^4$ to $10^6$~K (see e.g.~Dalgarno
\& McCray~1972). Since the immediate post--shock temperature for a strong
shock is given by
\begin{equation}
   T_{\rm s}={3 \over 16} {\mu m_{\rm H} \over k} v^2
\end{equation}
this means that shocks with $v$ between 30 and 400~km~s$^{-1}$ are likely to
be radiative. The velocity $v$ here is the pre--shock velocity in the frame
in which the shock is stationary.

As a star evolves through the post--AGB phase its wind velocity will go up
and mass loss rate down. This means that one expects the bubble to be
initially momentum--driven and later make a transition to
energy--driven. Kahn \& Breitschwerdt (1990) analytically calculated the
conditions for which this transition happens and found that for a wide range
of parameters the bubble makes the transition from momentum to energy--driven
when $v_{\rm fast} \sim 150$~km~s$^{-1}$. The reason for this is the strong
dependency of the post--shock cooling time on the wind velocity
\begin{equation}
  t_{\rm cool} = 0.255 {(1+\sqrt{\alpha})^2\over \alpha q}{v_{\rm slow}\over
  \dot{M}_{\rm slow}}v_{\rm fast}^5t^2\,,
\end{equation}
in which
\begin{equation}
  \alpha = {\dot{M}_{\rm fast}v_{\rm fast} \over \dot{M}_{\rm slow}v_{\rm
  slow}}\,,
\end{equation}
and $q$ is a constant from the assumed analytical cooling function which has
a value of $4\times 10^{32}$~cm$^6$~g$^{-1}$~s$^{-4}$ for normal cosmic
abundances.

Mellema~(1994) and Dwarkadas \& Balick~(1998) numerically calculated the
evolution of wind--driven bubbles with evolving fast winds and confirmed this
behaviour. Also when using detailed radiative cooling or a somewhat less
accurate cooling curve, the bubble makes the transition from momentum--driven
to energy--driven at a fast wind velocity of approximately 150~km~s$^{-1}$.

Dwarkadas \& Balick (1998) investigated the momentum--driven phase somewhat
more closely and found that during this phase the bubble is sensitive to the
Nonlinear Thin Shell Instability (NTSI). The effects of this instability
survive even beyond the momentum--driven phase as they found that bubbles
which evolved {\it through}\/ a momentum--driven phase possess a much more
disturbed interior.

Another property of bubbles in the momentum--driven phase is that they are
much more affected by asphericities in the fast wind. The reason for this is
that the ram pressure of the fast wind is the driving force, so if the wind
pushes harder in one direction than in another, the shape of the bubble will
reflect this. During the momentum--driven phase it is possible to form an
aspherical PNe using an aspherical fast wind and a spherical slow wind.

On the other hand, in the energy--driven phase, it is the thermal pressure of
the hot shocked fast wind which drives the bubble and as thermal pressure is
locally uniform, and any large scale pressure variations in the hot bubble
are quickly smoothed out, possible asphericities in the fast wind will during
the energy--driven phase only influence the shape of the bubble in a watered
down manner, if at all.

The `canonical' value of 150~km~s$^{-1}$ for the transition from one phase to
the other depends on
\begin{itemize}
\item Density of the fast wind, and hence its mass loss rate (since cooling
goes with $n^2$)
\item Details of the cooling process expressed by the value of $q$, which is
partly determined by the abundances.
\end{itemize}

Since the winds from [WR] stars have both high densities and peculiar
abundances one can expect the value of the transition velocity to change. To
find out how much requires a detailed calculation taking into account the
cooling rates of all the different ions, which as far as I know has never
been attempted. An approach similar to that described in Raga et
al.~(1997) should work well for this kind of problem.

A simple estimate can be made using Eq.~(6). Considering the
ionic cooling curves from Cox \& Tucker (1969) one can estimate that an
increase of the abundance of C by a factor 500-1000 (and factors 6 and 50 for
He and O respectively) will mean an increase in the cooling efficiency by a
similar factor. Raising $q$ by $10^3$ and $\alpha$ by $10^2$ in Eq.~(6) means
that the critical velocity goes up with a factor 10 or so. In other words up
to wind velocities of 1000~km~s$^{-1}$ or even higher the bubbles around [WR]
stars would be momentum--driven. This includes all late--type [WC] stars.

There are claims from bubbles around Pop.~I WR stars that this is in fact the
case (Chu 1982), but for the [WR] nebulae this question has as yet not been
considered. The consequences would be more chaotic nebulae due to the
Nonlinear Thin Shell Instability, and more extreme morphologies due to
aspherical mass loss during the post-AGB phase. 

For the first there is little indication in the observed morphologies,
although more careful analysis may show otherwise. There is the reported
need for a higher turbulent velocity needed to explain the line shapes from
PN around [WR] stars (Ge{\c s}icki \& Acker~1996), which may be due to this
effect.

As pointed out above, there is no evidence for the second effect. This
absence of any correlation between [WR] stars and nebular morphology then
implies that mass loss in the post--AGB phase does not deviate much from
spherical.

\section{Born--again Planetary Nebulae}

One model for the formation of [WR] stars is the occurrence of a late to very
late He--shell flash (Bl{\"o}cker, this volume). This model seems at the
moment to be able to explain the observed abundances the best. However, the
fact that there appear to be no differences between the nebulae around [WR]
and normal CSPN causes problems for this scenario.

The reason is that the formation of PN empties the region around the
star. The fast wind blows a bubble which is almost as empty as interstellar
space. Models show that a typical density inside a PN bubble is 1 to
10~cm$^{-3}$. If one assumes that a very late He-shell flash happens after
$10^4$~years and that the PN expands with a velocity of 30~km~s$^{-1}$, one
obtains a PN radius of $\sim 10^{16}$~cm at the time of the flash, which
means a total mass of about $10^{-8}$~M$_\odot$~yr$^{-1}$ interior to this
first PN, insufficient to sweep up a second PN, which would require between
0.01 and 1~M$_\odot$.

One can think of three possible solutions to this problem. Firstly, `reuse'
the old PN. The implication of this would be that the PNe around born--again
post--AGB stars are `old', at least older than the apparent age of the
star. In the case of A30, A78, and Sakurai's object this certainly seems to
be the case, but for the majority of [WR] stars this is not true, their PNe
are very similar to those seen around stars which supposedly evolved straight
off the AGB, implying that they have not suffered a late to very late
He-shell flash.

Perhaps it would be possible to accrete part of the old PN back to the
star. The scenario would be that when the fast wind stops, the pressure
inside the nebula starts dropping and material from the swept--up nebula
starts diffusing back in. There is little observational evidence for a
process like this, but it is true that the Helix nebula is actually not as
empty as one expects. The `hot bubble' is filled with cometary knots and a
more diffuse high ionization gas of a density of about 100~cm$^{-3}$ (see
e.g.\ Maeburn et al.~1998). The evolution of `old' PNe has not been studied
well and requires some more attention.  Still it is doubtful that a new PN
made out of the material of a diffused old PN would look the same as a
`normal' PN.

Thirdly, the star could lose 0.01 to 1~M$_\odot$ with a slow velocity, this
way mimicking AGB mass loss. Also this seems unlikely, since the stars do not
have that much mass to lose.

In all, the (very) late He--shell model seems to be irreconcilable with the
fact that the PNe around [WR] look so `normal'.

\section{Conclusions}

The [WR] phenomenon shows us again that it is most useful to consider a star
and its circumstellar environment together. The observed properties of the
nebula around these stars can help us in understanding the nature of the [WR]
phenomenon, and at the same time properties of [WR] stars can help us
understand the formation of PNe. To sum up the conclusions reached in this
paper:
\begin{enumerate}
\item The higher average expansion velocity of PNe around [WR] stars can be
understood as being due to the higher mass loss rates from the star.
\item The typical [WC] abundances are expected to lead to a longer lasting
momentum--driven phase in the formation of the nebulae. This phase may last
until the wind reaches velocities of 1000~km~s$^{-1}$.
\item A longer lasting momentum--driven phase should lead to the nebulae
becoming more affected by instabilities. This may be the explanation for the
fact that a higher turbulent velocity is needed to explain the line shapes of
PNe around [WR] stars.
\item A longer lasting momentum--driven phase allows asphericities in the
stellar wind to have a larger effect on the shape of the PN. Since the
observed nebulae show no sign of this, it implies that the post--AGB wind is
mostly spherical.
\item Producing a second PNe in the case of a born--again PN scenario is
difficult to impossible. Consequently, born--again PNe should show a
discrepancy between the age of the PN and that of the star. Since this is not
observed for most [WR] stars, the born--again scenario seems not to apply.
\item The lack of correlation between PN morphology and the [WR] phenomenon
shows that the ultimate mechanism to produce aspherical PNe is unrelated to
the process which produces the [WR] star.
\end{enumerate}

\end{document}